\documentclass[journal=nalefd,manuscript=communication]{achemso}
\usepackage[version=3]{mhchem} 

\author{Jeil Jung}
\affiliation{Department of Physics, The University of Texas at
Austin, Austin, Texas 78712, USA}
\author{Zhenhua Qiao}
\affiliation{Department of Physics, The University of Texas at
Austin, Austin, Texas 78712, USA}\email{zhqiao@physics.utexas.edu}
\author{Qian Niu}
\affiliation{Department of Physics, The University of Texas at
Austin, Austin, Texas 78712, USA}
\alsoaffiliation{International Center for Quantum Materials, Peking University, Beijing 100871, China}
\author{Allan H. MacDonald}
\affiliation{Department of Physics, The University of Texas at
Austin, Austin, Texas 78712, USA}

\title{Transport Properties of Graphene Nanoroads in Boron-Nitride Sheets}
\begin{document}
\begin{abstract}
We demonstrate that the one-dimensional (1D) transport channels that appear in the gap when graphene nanoroads are embedded in boron-nitride (BN) sheets are more robust when they are inserted at AB/BA grain boundaries. Our conclusions are based on {\em ab-initio} electronic structure calculations for a variety of different crystal orientations and bonding arrangements at the BN/C interfaces. This property is related to the valley-Hall conductivity
present in the BN band structure and to the topologically protected kink states that appear in continuum Dirac models with position dependent masses.
\end{abstract}
Metallic transport channels appear at the edges, surfaces,
and interfaces of two and three
dimensional bulk insulators when a bulk topological index
changes value as the interface region is crossed.
\cite{polymer,volovik,semenoff,yaowang,morpurgo,multilayer,highway,arun1,arun2,arun3}
This property can provide transport channels
in otherwise insulating materials.
The metallic states possess an internal structure
related to their sense of propagation which leads to
special transport properties including zero bend resistance
at sharp turns in the current propagation trajectory, pseudospin memory,
and suppressed backscattering. \cite{highway}

\begin{figure}
\includegraphics[width=8.5cm,angle=0]{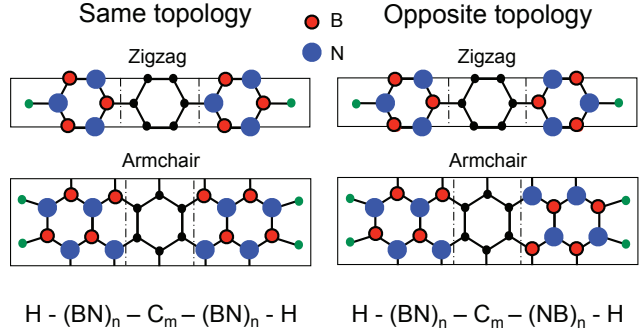} \\
\caption{(Color online) Schematic plot of two types of zigzag and
armchair BNC nanoroads.
The integer $n$ is the number of atoms in each BN partition whereas
$m$ is the number of C atoms across the nanoroad.
The outermost BN atoms are passivated with hydrogen.
{\em Left Panel:} The {\em same topology} arrangement in which the sign of the
valley Hall effect is the same in both BN sheets. In the absence of C atoms the
BN fragments can be joined seamlessly.
{\em Right Panel:} The {\em opposite topology} arrangement in which the sign of the
valley Hall effect is opposite in the two BN sheets. In the absence of C atoms
it is necessary to form a nearest-neighbor bond between atoms of the same species-
either N$-$N and/or B$-$B.
} \label{setup}
\end{figure}

\begin{figure}
\includegraphics[width=14cm,angle=0]{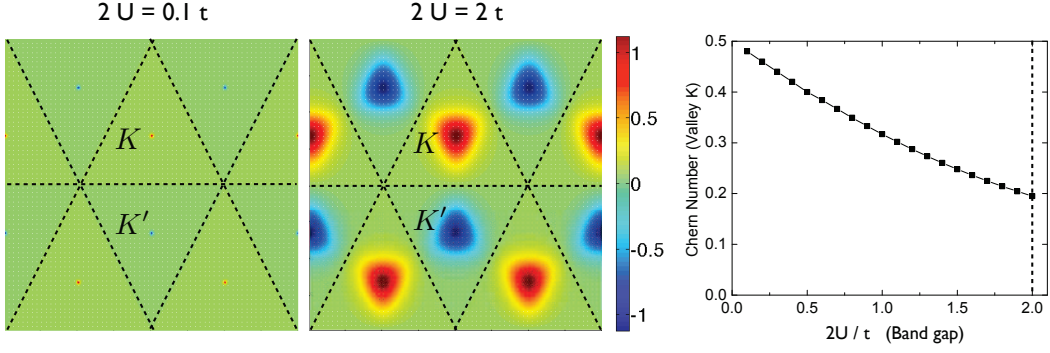} \\
\caption{
Berry curvatures obtained from a tight-binding model of the honeycomb lattice
with different band gap sizes $\Delta = 2 U$ where $U$ is the
staggering site potential. The Berry curvatures are more sharply peaked near the Brillouin zone
corners $K$ and $K'$ when the band gap is small and they spread out more
when the gap is larger.
The rightmost panel shows the deviation of the valley resolved Chern number from the ideal 1/2
value of the continuum model obtained integrating the Berry curvature in one equilateral
triangle corresponding to
one half of the primitive cell associated to valley $K$ given by $C_K = (2 \pi)^{-1}  \int_{K} d^2 k \Omega({\bf k})$.
The dotted vertical line at $2U/t = 2$ indicates the valley resolved Chern number obtained for the tight-binding
hamiltonian we used to approximate the BN bands.
} \label{berry}
\end{figure}

In this Letter we show that two-dimensional hybrid structures
consisting of graphene nanoroads \cite{hybrid} (See \ref{setup}) embedded
in hexagonal boron nitride (BN) sheets can be an attractive host for
topologically assisted one-dimensional (1D)
transport channels.  Our study is motivated by the observation that the $\pi$-bands of BN are similar to those
of a graphene sheet, except that the honeycomb sublattices have different $\pi$-electron site energies.
(The electrostatic potential is more attractive on the higher-Z
N atom sites).  Because valence band states near the $K$ and $K'$ Brillouin-zone corners are
strongly localized on $N$ sites whereas those elsewhere in the Brillouin-zone are divided more
evenly, the bands of BN contribute with valley Hall conductivities  
and associated Berry curvatures
of opposite signs near $K$ and $K'$ whose continuum model expression is given by
$\Omega({\bf q}) = \tau_z  \left( 3a^2 \Delta t^2 \right) /  \left( 2 \left( \Delta^2 + 3q^2 a^2 t^2 \right)^{3/2} \right)$
where ${\bf q}$ is the momentum measured from a Dirac point, $\tau_z = \pm 1$ is the valley, $a$ is the lattice
constant of the honeycomb lattice and $\Delta$ is the band gap.
\cite{xiao}.
If the B-site {\em vs.} N-site potential difference was small, these Berry curvatures would
be strongly concentrated near the BZ corners [see \ref{berry}].
Indeed, in the graphene case it is known that electronic properties in
systems with sublattice-staggered potentials that
are weak compared to the $\pi$-band width can be described
using a two-dimensional massive Dirac equation.
Under these circumstances
Dirac-equation continuum models are valid and predict one-dimensional (1D) localized
states along lines where the sublattice-staggered potential (the Dirac-equation mass)
changes sign.\cite{semenoff,yaowang}
In the closely related bilayer\cite{morpurgo,multilayer,highway} and multilayer\cite{multilayer}
graphene cases, the sublattice staggered potentials
are readily generated experimentally by applying an electric field across the layers and
varied spatially by appropriately arranging external gate voltages.
In the single-layer case, however, it has not been obvious how the
sublattice-staggered potentials could be realized,
although gaps may be present with a lattice matched BN substrate \cite{bngap} or when 
gate potential profiles are carefully correlated with strains \cite{low}.
This work is motivated by the idea that the difference between B-atom and N-atom potentials in BN provides
the desired staggered potential.  Of course the staggered potential is not weak in the
case of a BN crystal, so that expectations based on massive Dirac continuum models
must be checked by {\em ab initio} electronic structure calculations.

\begin{figure*}[t]
\includegraphics[width=17cm,angle=0]{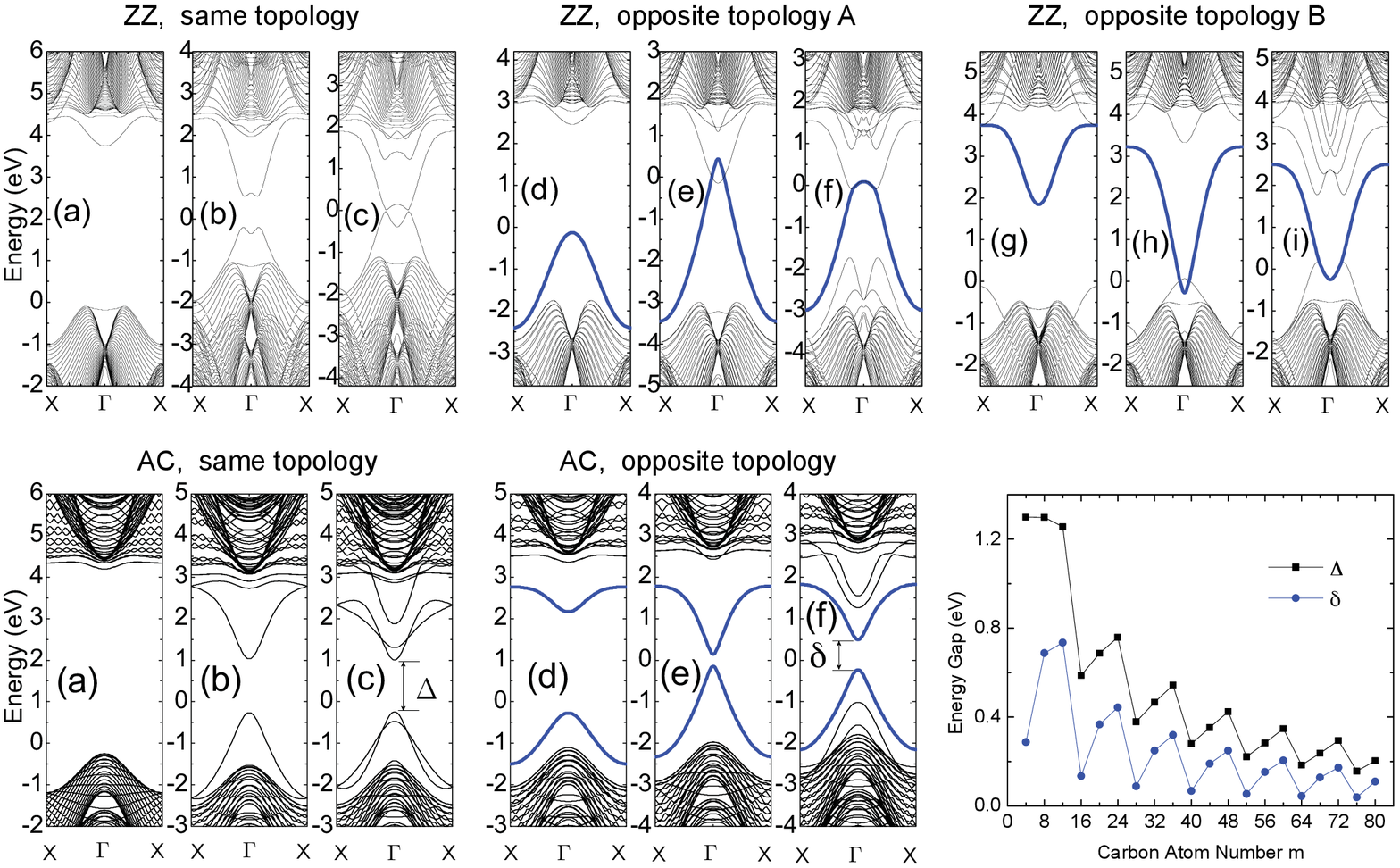}
\caption{(Color online)
{\em Upper Panel:}
Quasi 1D band structures of zigzag (ZZ) graphene nanoroads of different widths embedded
in hexagonal boron nitride.  The three band structures are for widths corresponding to
zero, one and three complete hexagons or equivalently $m$ = 0, 4 and 12 carbon atoms in a unit cell.
The BN width number $n=40$ for all cases.
We consider the {\em same topology} configuration, (BN)$_n -$ C$_m -$ (BN)$_n$,
the  {\em reversed topology} configuration A, (BN)$_n -$ C$_m -$ (NB)$_n$,
in which nitrogens bind with carbon and
reversed topology configuration B, (NB)$_n -$ C$_m -$ (BN)$_n$,
in which borons bind with carbon.
States that are peaked in the middle of the nanoroads are represented by thick blue lines.
{\em Lower Panel:}
Band structures of armchair (AC) graphene nanoroads of different widths embedded
in hexagonal boron nitride.  The two band structures are for
$m$ = 0, 4 and 12 carbon atoms in a unit cell, corresponding to nanoroads that are
$0$, $1$ and $3$ hexagons wide.
Same and reversed topology configurations are considered.
The states that are localized in the middle of the nanoroads are again
indicated by thick blue lines.  As in the case of graphene nanoribbons the
gaps have an oscillatory behavior with period $12$ (corresponding to 3 hexagons),
embedded in a smooth envelope that follows a $m^{-1}$ law.}
\label{armchair_nanoroads}
\end{figure*}

The topological index that is inherent in BN
bands can be viewed as a valley Hall effect since each valley separately supports
half integer quantum Hall effects of opposite sign, or at least does in the limit that the potential
difference is small compared to the $\pi$-band width.  The properties that we discuss below
are therefore closely related to the simple momentum space Berry curvature pattern
in the bands of BN sheets.
Hybrid BN/graphene systems of the type we study occur
naturally in patched sheets containing a mixture of
atomically thin graphene and BN, \cite{rice}
and have recently become a subject of great interest,
\cite{rubio,hybrid,gaparm1,bndoping,pruneda,dotsinbn,colorful,embribbon,halfmet,birribon,
mixture,hybrid_stab,postsynthesis,af_island,boronitrene,hybrid_stab,
transport,transport_armch,transport_hybr,had_graphenebn}
as the basis of a possible strategy for controlling graphene system band gaps.
In this work we show that the properties of the transport channels formed
by graphene nanoroads in BN strongly depend on the valley Hall effect of the surrounding material.


The simplest graphene nanoroad geometries
have C/BN interfaces with crystallographic orientations
along the zigzag and armchair directions.
The unit cells considered in our {\em ab initio}
calculation are illustrated in \ref{setup}.
In each case a graphene nanoribbon is flanked on the left and right by
BN sheets with B and N atoms either on the same or opposite sublattices.
When the two different BN fragments can be joined seamlessly preserving
the crystalline B and N atom sequences upon carbon atom removal,
as in the cases illustrated in the left panels of \ref{setup}, the BN sheets on left
and right have valley Hall conductivities of the same sign and we
will say that they have the {\em same topology}.
The right and center panels show junctions in which the BN sheets on opposite
sides of the graphene nanoroad have opposite
valley Hall conductivities.  We will refer to these configurations as having {\em opposite topologies}.
When the carbon atoms are removed,
joining the BN fragments would in this case require B$-$B or N$-$N bond.
As shown in \ref{armchair_nanoroads}, a difference in the topology of the BN sheet arrangement
on the nanoroad shoulders invariably leads to a qualitative change in
nanoroad quasi-1D band structure.

The indices $n$ and $m$ in \ref{armchair_nanoroads} represent respectively the number of BN and carbon
atoms in the 1D unit cell.
The width of the BN fragment, specified
by the index $n$, plays a minor role in mid-gap state properties.
We begin by discussing results for zero, one, and three hexagon width
zigzag nanoroads presented in the upper panels of
\ref{armchair_nanoroads}.
The zero width cases illustrates the electronic structure of
seamless BN nanoribbons with $2n = 80$ atoms of BN, and
therefore a width of 86 nm.  The large band gap of 4 $eV$
is comparable to that of bulk BN.
There is a notable lack of particle-hole symmetry in the band structure,
which is partly due to the difference between B and N site potentials mentioned previously
but cannot be captured by a nearest neighbor tight-binding model.
We are mainly interested in the mid-gap states which emerge when a graphene nanoroad
is embedded in the BN sheet.  As expected on the basis of the properties of
pure graphene nanoribbons, we find that zigzag graphene nanoroads do
have strongly reduced gaps because of the presence of zigzag
edge states.\cite{fujita,nakada,superexchange} There is still however a topology sensitive feature;
in the opposite topology case
a mid-gap state appears that is peaked near the center of the nanoroad,
rather than at its edges, with some penetration into the BN region.
This state is indicated by blue shading in Figs. 3(d) to  3(i).
We view this state as the lattice remnant of the mass-reversal (kink) state which would appear in the
opposite topology case if the staggered potential was weak and the continuum model applied.
Even in the absence of carbon atoms its appearance reduces the energy gap by approximately 2 eV.
If the continuum model was accurate there would be opposite velocity kink states associated with the
$K$ and $K'$ valleys and the gap would disappear entirely.

Although gaps are absent for wide
nanoroads in both like cases, the additional kink state present for opposite topologies closes the
gap already at one hexagon width in the opposite topology case.
For same topology configurations, the gaps are closed because the
$k_x \sim 0$ states that are most strongly
localized at the edge are shifted in opposite directions by coupling to the
BN shoulders. \cite{superexchange}
For inverted topology,  since the same type of atoms N or B are attached at both edges of the
graphene nanoroad, we see a shift of onsite energies at the edge sites in the same direction,
upwards for N atom bonds and downwards for B atom bonds.
In this case the states closing the band gap have kink-state character and
are spread widely over the C atoms of the nanoroad,
and with some penetration into the BN region. \cite{highway}

We now turn to armchair nanoroads.  In ribbons this orientation does not
support metallic edge states \cite{breyfertig}, and instead yields gaps which
scale as the inverse ribbon width.  Continuum model considerations suggest
that reversed topology would lead to vanishing gaps for all widths.\cite{multilayer,highway}
The band structures in the lower panel of
\ref{armchair_nanoroads} demonstrate that there are gaps in the reversed
topology case, but that the gaps are much smaller than in the same topology case.
The magnitude of the small avoided crossing gap between opposite velocity
kink states at $k_x \sim 0$ shows the same oscillatory decline as a function of ribbon width
that has been heavily studied in ribbons.\cite{son,breyfertig}
The gap reaches a value as large as 0.75 eV for armchair single-hexagon nanoroads.
It is noteworthy that this avoided crossing is always smaller than the finite-size gap in
either nanoribbons of the same width or same topology nanoroads,
a fact that allows the kink channels to become the main current
conducting path when the Femi level of the system is adequately shifted. 
The same topology and opposite topology gaps, $\Delta$ and $\delta$,
are plotted as a function of nanoroad width
in the bottom right panel of \ref{armchair_nanoroads}.

Zigzag graphene nanoroads have bearded interfaces
terminations when they have an odd number of carbon atoms or non modulo 4 even integer number
of C atoms in the unit cell.\cite{fujita,nakada}
These interfaces  are not expected to be common in experimental
systems because of stability considerations.
A combination of zigzag and armchair fragments \cite{dresselhaus} are more
abundant than bearded configurations with repeated dangling atoms.
Bearded edges lead to more complicated band structures that we discuss
further in the supplementary information,
but generally preserve the tendency toward greater conduction in the
opposite topology case.

\begin{figure*}[t]
\includegraphics[width=8cm]{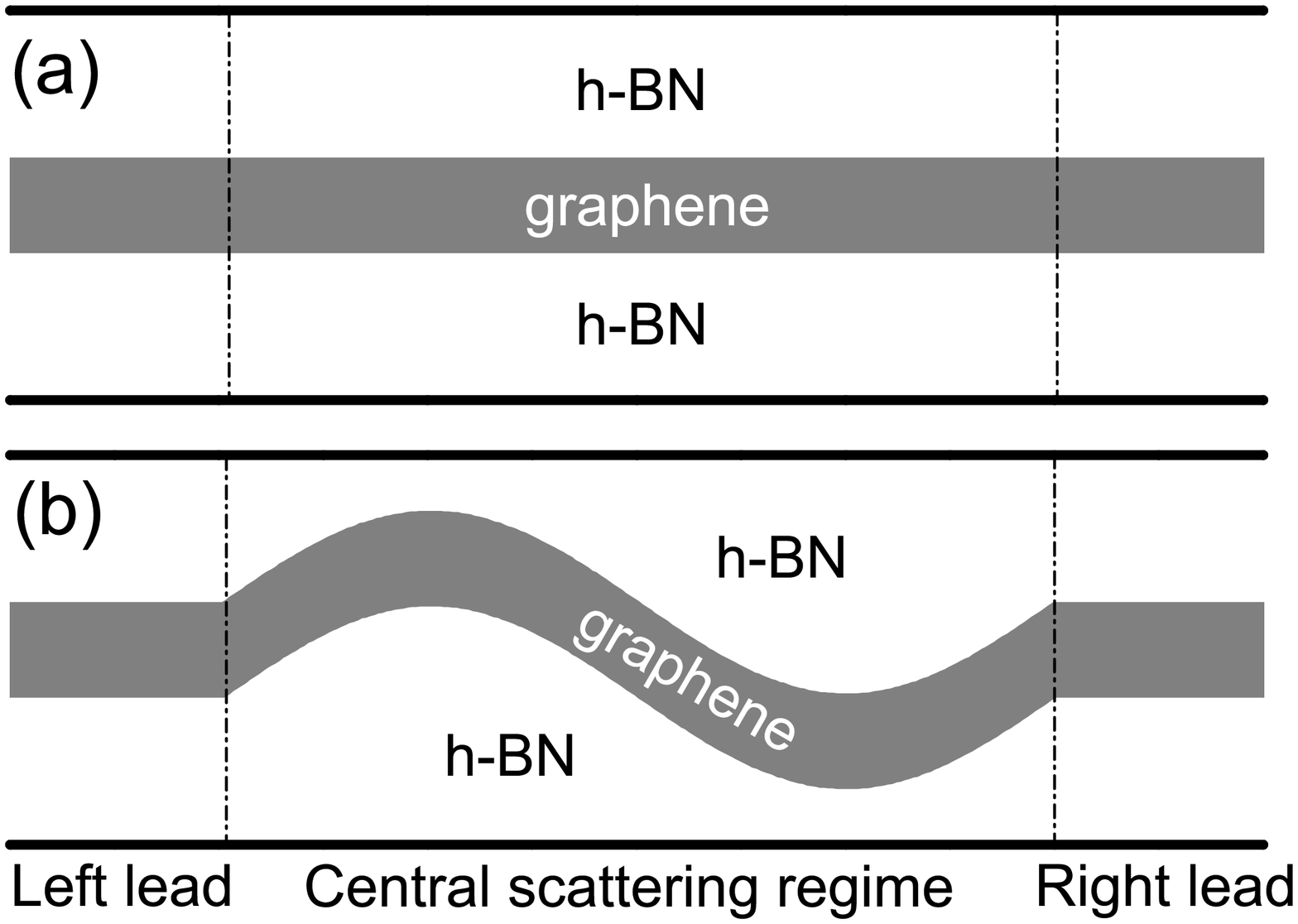}
\caption{Schematic plot of the two-terminal BNC nanoroad transport measurement that we model.
(a) Straight BNC nanoroads; (b) Curved BNC nanoroads: the
two leads are exactly the same as those in panel (a).
The nanoroad will support kink-like states when the two BN regions have opposite topology.}
\label{transportsetup}
\end{figure*}

To address the influence of topology on nanoroad transport,
we consider a two-terminal geometry with a central scattering region containing nanoroads that
are either straight, or curved with both armchair and zigzag segments
[See \ref{transportsetup}].
In our model, the source and drain electrodes are semi-infinite zigzag graphene nanoroads.
The central scattering region contains 96$\times$60 lattice sites and
the nanoroad width is fixed at 32 carbon atoms.
The BNC nanoroad system is modeled using a nearest neighbor tight-binding Hamiltonian of the form
\begin{eqnarray}
H =-t\sum_{\langle ij \rangle} ~ c_i^{\dag} c_j +
\sum_{i \in A} U_{A} c_{i}^{\dag} c_{i}-\sum_{i \in B} U_{B} c_{i}^{\dag} c_{i}, \label{eq1}
\end{eqnarray}
where $c^{\dag}_i$~($c_{i}$) is a $\pi$-orbital creation~(annihilation) operator for an electron at the
site $i$, $U_{A}$ and $U_{B}$ are the BN $\pi$-orbital site energies on the $A$ and $B$ subllices,
and $t=2.6~eV$ is the nearest neighbor hopping energy for both graphene and h-BN.
We introduce a staggered \emph{AB} sublattice potential to describe the site energy differences between
B and N sublattices of h-BN, {\em i.e.} $U_B=-U_N=\lambda U_0$.
Here, $\lambda$=`$\pm$' defines the crystal topology of the h-BN regions in the graphene nanoroad.
$U_0$ measures the strength of the staggered potential, and for the purposes of this illustrative
calculation we have assumed it to be $U_0 = t$.

The temperature $T=0$ conductance from the left lead to the right lead can be calculated using
 the multi-probe Landauer-B\"{u}ttiker formula:~\cite{datta}
\begin{equation}
G_{RL}=\frac{e^2}{h}~{\rm Tr} [\Gamma_{R} {\rm G}^r \Gamma_{L} {\rm G}^a],
\end{equation}
where ${\rm G}^{r,a}$ are the retarded and advanced Green functions of the central scattering region.
The quantity $\Gamma_{R/L}$ is a linewidth function describing the coupling between left/right-lead
and the central region and is obtained by calculating the self-energy $\Sigma^{r}_{R/L}$ of
left/right-lead using a variant transfer matrix method. \cite{transfer,varianttransfer,JianHong}

\begin{figure}[t]
\includegraphics[width=8cm]{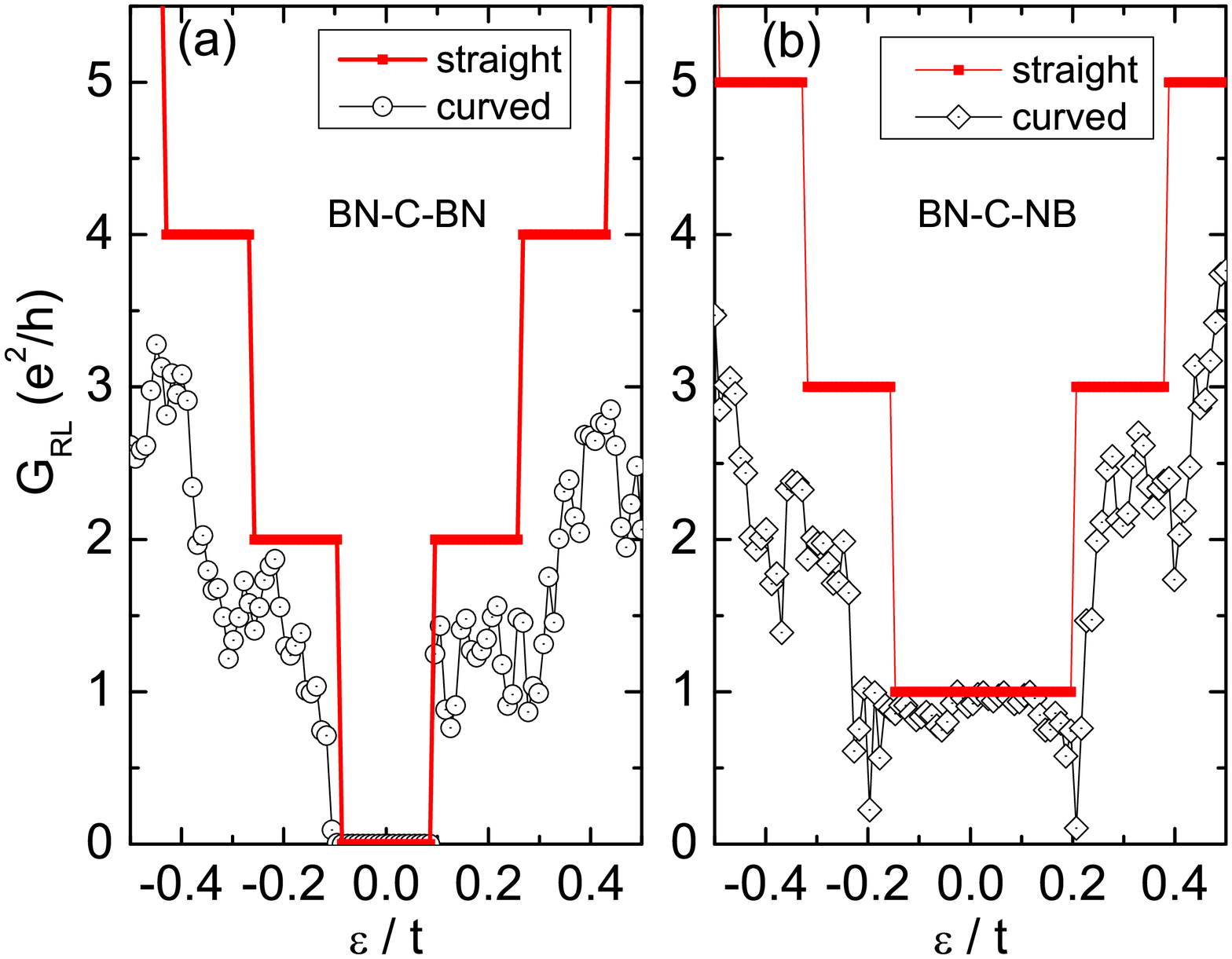}
\caption{(Color online)
Conductance calculations for straight zigzag-edged and curved
graphene nanoroads embedded in the h-BN as a function of the Fermi energy $\varepsilon$.
Panel (a): conductance for graphene nanoroads embedded within h-BN sheets with the same
topology. An insulating gap clearly arises near the charge neutrality point.
Panel (b): graphene nanoroads embedded within h-BN sheets with different
topologies.  The conductance plateau near the charge neutrality region is due to
a topologically supported edge state.  Irregular and curved edges introduce
a extremely small backscattering probability.} \label{conductance}
\end{figure}

\ref{conductance} plots conductance $G_{RL}$ as a function of Fermi energy $\varepsilon$ for the straight (zigzag-edged) and curved BNC nanoroads. In panel (a), the crystal topologies are the same in upper and lower h-BN regions.
We observe that the conductance of the straight BNC nanoroad is quantized to be $G=2n~e^2/h$ ($n=0,1,2...$).
The conductance plotted here is per spin, so that the even conductance values can be attributed to the
presence of ballistic states that are localized at both edges.
The nanoribbon finite size effect for this nanoroad width produces
a small band gap with vanishing conductance near the charge neutrality point.
When the graphene nanoroad becomes curved in the central regime, disorder
produces backscattering which supresses the conductance below the quantized value.

The case of opposite BN shoulder topologies 
is illustrated in \ref{conductance}(b).
In this case the conductance rule for straight BNC nanoroads is changed to $G= \left(2n-1 \right)~e^2/h$ ($n=1,2...$) as shown in \ref{conductance}(b).  This change reflects the addition of a topologically supported
kink state.  For the curved reversed topology nanoribbon, the conductance
does not decrease substantially in the region near the charge neutrality point
where the only remaining channel is the kink state.
Near the neutrality point the conductances show a surprising plateau feature with minor fluctuations that
may be suppressed further by increasing the width of the graphene nanoroad.
This finding strongly reflects the robustness of the crystal topology induced 1D topological conducting state against BNC nanoroad bends and is consistent with the finding of the zero bend resistance for
1D topological confinement states in gated bilayer graphene.\cite{highway}

\begin{figure}[t]
\includegraphics[width=8cm]{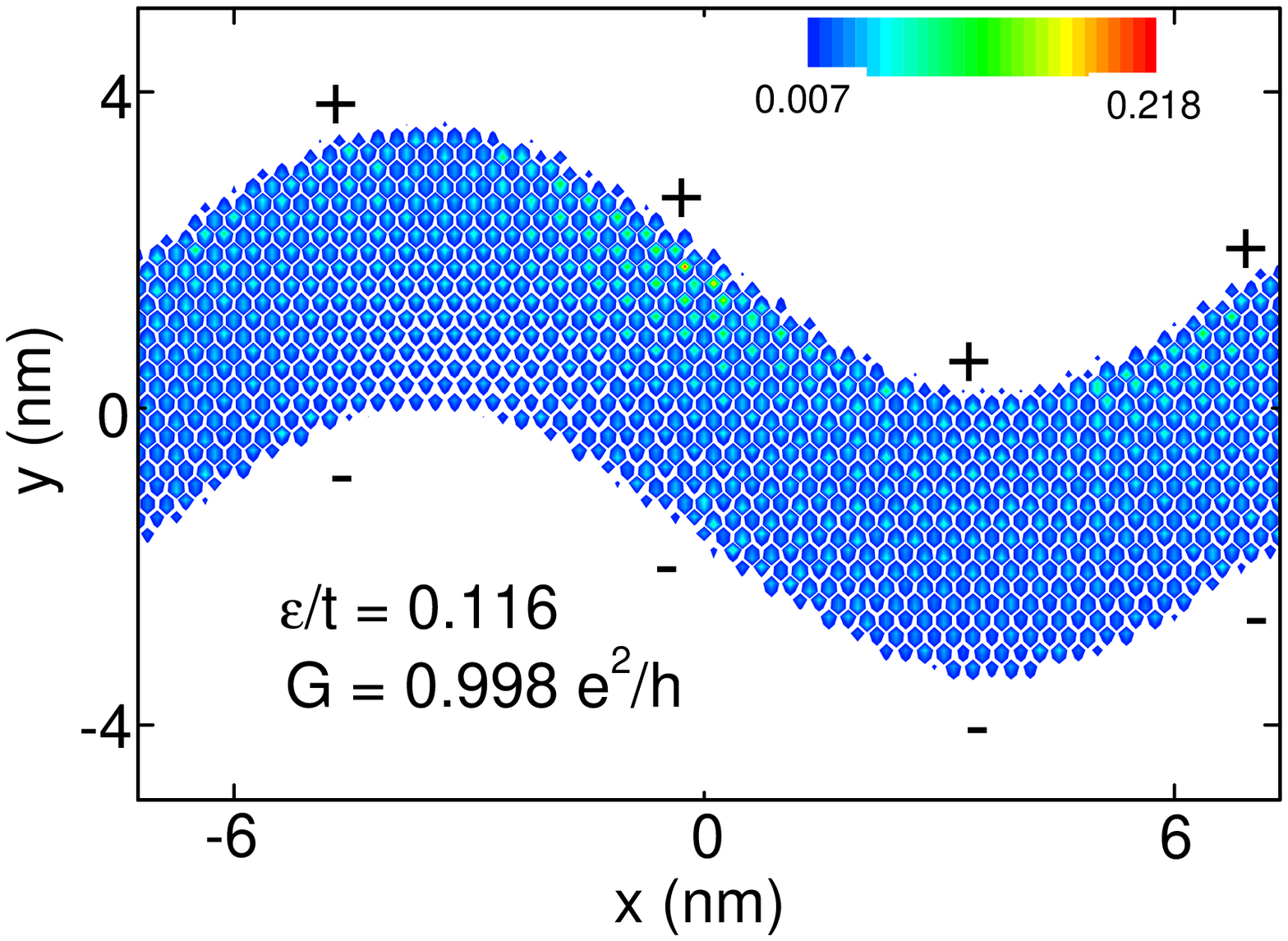}
\caption{(Color online)
Local density-of-states of a kink state in a curved carbon nanoroad embedded
between opposite topology BN sheets, suggested here by the `$+$' and `$-$' signs,
calculated through a nearest neighbor tight-binding model as discussed in the main text.
The edges consist of a mixture of zigzag and armchair fragments but the
disorder does not degrade transmission.
} \label{ldos}
\end{figure}

To visualize the 1D topological conducting channel, we
can evaluate the position dependence of its local density of states (LDOS) of the current originating
from $P^{th}$ lead using
\begin{equation}
\rho_{\rm p}({\bf r}, \varepsilon)=\frac{1}{2\pi}[{\rm G}^r \Gamma _{\rm p}
{\rm G}^a]_{{\bf r} {\bf r}}.
\end{equation}
Here $\varepsilon$ is the Fermi energy. In \ref{ldos}, we plot the LDOS distribution of the
topologically assisted conducting channel from the left lead into a snake-shaped BNC nanoroad.
Our calculation shows that it is localized across the graphene atomic sites
with only a small portion on the interface between h-BN and graphene.
The transport calculations for the straight and curved nanoroad geometries
confirms that the kink states picture remains essentially
valid even when both zigzag and armchair edges
are simultaneously present.  The almost ballistic transport is
preserved in spite of irregular bonds
at the graphene/BN interfaces.

In summary, we have shown that graphene nanoroads embedded in BN sheets
host kink-like transport channels when surrounded by opposite topology BN fragments.
Transmission by the kink-states is topologically assisted and nearly perfect
even when the graphene nanoroad is curved and has irregular edges consisting
of combinations of both for zigzag and armchair edges.

\section{Calculation Details}
The self-consistent {\em ab-initio} calculations were carried out using the
PAW-LDA parametrization of the exchange correlation functional in a plane wave basis code,
as implemented in Quantum Espresso. \cite{pwscf}
We used the 1.443 Angstrom 
for the separation between neighboring B, N and C atoms
and 1.09 $\AA$ for B$-$H and N$-$H separations.
These values are consistent with optimized lattice constants obtained from total energy minimization.
Relaxation to obtain a final crystal geometry with atomic forces below
$0.01 ~eV /$Angstrom   
had a negligible effect in modifying the interatomic distances and band structures.
For our nanoroad calculations in ribbon geometries we used 30, 1 and 1
k-points in the $k_x$, $k_y$ and $k_z$ directions, and a plane wave energy cutoff of 60 ~Ry.
The vacuum separation distances between supercell repeated nano ribbons
was set to $20 \AA$ in both $\hat{y}$ and $\hat{z}$ directions.

\begin{acknowledgement}
This work was supported by NSF (DMR 0906025), NRI-SWAN, Welch Foundation (TBF1474, F-1255), DOE (DE-FG03-02ER45958, Division of Materials Science and Engineering), and Texas Advanced Research Program. We gratefully acknowledge the computation resources and assistance provided by the Texas Advanced Computing Center.
\end{acknowledgement}
\begin{suppinfo}
We have also attached a supporting information material for a more
detailed understanding of our work.
\end{suppinfo}


\begin{thebibliography}{99}




\bibitem{polymer}
Heeger A. J.; Kivelson S.; Schrieffer J. R.; Su W.-P. \textit{Rev. Mod. Phys.} 1988, \textbf{60}, 781.

\bibitem{volovik}
Volovik G. E., `The Universe in a Helium Droplet', Oxford University Press (2003).

\bibitem{semenoff}
Semenoff G. W.; Semenoff V.; Zhou F. \textit{Phys. Rev. Lett.} 2008, \textbf{101}, 087204.

\bibitem{yaowang}
Yao W.; Yang S. A.; Niu Q. \textit{Phys. Rev. Lett.} 2009, \textbf{102}, 096801.

\bibitem{morpurgo}
Martin I.; Blanter Ya. M.; Morpurgo A. F. \textit{Phys. Rev. Lett.} 2008, \textbf{100}, 036804.

\bibitem{multilayer}
Jung J.; Zhang F.; Qiao Z. H.; MacDonald A. H. \textit{Phys. Rev. B} 2011, \textbf{84}, 075418.


\bibitem{highway}
Qiao Z. H.; Jung J.; Niu Q.; MacDonald A. H. \textit{Nano Lett.} 2011, \textbf{11} (8), 3453.

\bibitem{arun1}
Killi M.; Wei T.-C.; Affleck I.; Paramekanti A. \textit{Phys. Rev. Lett.} 2010, \textbf{104}, 216406.

\bibitem{arun2}
Killi M.; Wu S.; Paramekanti A. \textit{Phys. Rev. Lett.} 2011, \textbf{107}, 086801.

\bibitem{arun3}
Wu S.; Killi M.;  Paramekanti A.  E-print arXiv:1202.1441 (2012).

\bibitem{bngap}
Giovannetti G.; Khomyakov P. A.; Brocks G.;  Kelly P. J.;  van den Brink J.;
Physl Rev. B 2007, \textbf{76}, 073103.

\bibitem{low}
Low T., Guinea F., and Katsnelson M. I.,  \textit{Phys. Rev. B} 2011, \textbf{83}, 195436.




\bibitem{hybrid}
Bhowmick S.; Singh A. K.; Yakobson B. I. \textit{J. Phys. Chem. C} 2011, \textbf{115}, 9889.


\bibitem{xiao}
Xiao D.;  Yao W.; and  Niu Q., \textit{Phys. Rev. Lett.} 2007, \textbf{99}, 236809.



\bibitem{rice}
Ci L.; Song L.; Jin C.; Jariwala D.; Wu D.; Li Y.; Srivastava A.; Wang Z. F.; Storr K.; Balicas L.; Liu F.; Ajayan P. M. \textit{Nat. Mat.} 2010, \textbf{9}, 430.

\bibitem{rubio}
Rubio A. \textit{Nat. Mat.} 2010, \textbf{9}, 379.


\bibitem{gaparm1}
Seol G.; Guo J. \textit{App. Phys. Lett.} 2011, \textbf{98}, 143107.

\bibitem{bndoping}
Shinde P. P.; Kumar V. \textit{Phys. Rev. B} 2011, \textbf{84}, 125401.

\bibitem{pruneda}
Pruneda J. M. \textit{Phys. Rev. B} 2010, \textbf{81}, 161409(R).

\bibitem{dotsinbn}
Li J.; Shenoy V. B. \textit{Appl. Phys. Lett.} 2011, \textbf{98}, 013105.

\bibitem{colorful}
Liu Y.; Bhowmick S.; Yakobson B. I. \textit{Nano Lett.} 2011, \textbf{11}, 3113.

\bibitem{embribbon}
Ding Y.; Wang Y.; Ni J. \textit{Appl. Phys. Lett.} 2009, \textbf{95}, 123105.

\bibitem{halfmet}
Liu Y.; Wu X.; Zhao Y.; Zeng X. C.; Yang J. \textit{J. Phys. Chem. C} 2011, \textbf{115}, 9442.

\bibitem{birribon}
Fan Y.; Zhao M.; Zhang X.; Wang Z.; He T.; Xia H.; Liu X. \textit{J. Appl. Phys.} 2011, \textbf{110}, 034314.

\bibitem{mixture}
Lam K.-T.; Lu Y.; Feng Y. P.; Liang G. \textit{Appl. Phys. Lett.} 2011, \textbf{98}, 022101.

\bibitem{postsynthesis}
Berseneva N.; Krasheninnikov A. V.; Nieminen R. M. \textit{Phys. Rev. Lett.} 2011, \textbf{107}, 035501.

\bibitem{af_island}
Ramasubramaniam A.; Naveh D. \textit{Phys. Rev. B} 2011, \textbf{84}, 075405.

\bibitem{boronitrene}
Obodo K. O.; Andrew R. C.; Chetty N. \textit{Phys. Rev. B} 2011, \textbf{84}, 155308.

\bibitem{hybrid_stab}
Manna A. K.; Pati S. K. \textit{J. Phys. Chem. C} 2011, \textbf{115}, 10842.





\bibitem{transport}
Song L. \textit{et al.}, E-print arXiv:1105.1876 (2011).

\bibitem{transport_armch}
Modarresi M.; Roknabadi M. R.; Shahtahmasbi N. \textit{Physica E} 2011, \textbf{43} (9), 1751.

\bibitem{transport_hybr}
Qiu M.; Liew K. M. \textit{J. Appl. Phys.} 2011, 110, 064319.

\bibitem{had_graphenebn}
Cao T.; Feng J.; Wang E. G. \textit{Phys. Rev. B} 2011, \textbf{84}, 205447.

\bibitem{bearded}
Wakabayashi K. \textit{Phys. Rev. B} 1999, \textbf{59}, 8271.

\bibitem{dresselhaus}
Jia X. \textit{et al.} \textit{Science} 2009, \textbf{323}, 1701.

\bibitem{fujita}
Fujita M.; Wakabayashi K.; Nakada K.; Kusakabe K. \textit{J. Phys. Soc. Jpn.} 1996, \textbf{65}, 1920.

\bibitem{nakada}
Nakada K.; Fujita M.; Dresselhaus G.; Dresselhaus M. S. \textit{Phys. Rev. B} 1996, \textbf{54}, 17954.

\bibitem{superexchange}
Jung J.; Pereg-Barnea T.; MacDonald A. H. \textit{Phys. Rev. Lett.} 2009, \textbf{102}, 227205.

\bibitem{son}
Son Y.-W.; Cohen M. L.; Louie S. G. \textit{Phys. Rev. Lett.} 2006, \textbf{97}, 216803.

\bibitem{breyfertig}
Brey L.; Fertig H. A. \textit{Phys. Rev. B} 2006, \textbf{73}, 235411.

\bibitem{datta}
Datta S., `Electronic Transport in Mesoscopic Systems', Cambridge University Press (1995).

\bibitem{transfer}
L\'{o}pez Sancho M. P.; L\'{o}pez Sancho J. M.; Rubio J. \textit{J. Phys. F: Met. Phys.} 1984, \textbf{14}, 1205.

\bibitem{varianttransfer}
Qiao Z. H.; Wang J. \textit{Nanotech.} 2007, \textbf{18}, 435402.

\bibitem{JianHong}
Wang J.; Guo H. \textit{Phys. Rev. B} 2009, \textbf{79}, 045119.

\bibitem{pwscf}
Giannozzi P. {\em et al.}, http://www.quantum-espresso.org.
\end{thebibliography}

\end{document}